\documentclass[final,authoryear,3p,times,twocolumn]{elsarticle}
\usepackage{graphics}
\usepackage{amssymb}

\journal{Journal of Theoretical Biology}

\begin{document}
\begin{frontmatter}

\title{Coveting thy neighbors fitness as a means to resolve social dilemmas}

\author[Zhen]{Zhen Wang\corref{zw}}
\cortext[zw]{zhenwang@mail.nankai.edu.cn}
\author[Perc]{Aleksandra Murks}
\author[Du]{Wen-Bo Du}
\author[Rong]{Zhi-Hai Rong}
\author[Perc]{Matja{\v z} Perc\corref{mp}}
\cortext[mp]{matjaz.perc@uni-mb.si}

\address[Zhen]{School of Physics, Nankai University, Tianjin 300071, China}
\address[Perc]{Department of Physics, Faculty of Natural Sciences and Mathematics, University of Maribor, Koro{\v s}ka cesta 160, SI-2000 Maribor, Slovenia}
\address[Du]{School of Computer Science and Technology, University of Science and Technology of China, Hefei 230026, China \& \\ School of Electronic and Information Engineering, Beihang University, Beijing 100083, China}
\address[Rong]{Department of Automation, Donghua University, Shanghai 201620, China}

\begin{abstract}
In spatial evolutionary games the fitness of each individual is traditionally determined by the payoffs it obtains upon playing the game with its neighbors. Since defection yields the highest individual benefits, the outlook for cooperators is gloomy. While network reciprocity promotes collaborative efforts, chances of averting the impending social decline are slim if the temptation to defect is strong. It is therefore of interest to identify viable mechanisms that provide additional support for the evolution of cooperation. Inspired by the fact that the environment may be just as important as inheritance for individual development, we introduce a simple switch that allows a player to either keep its original payoff or use the average payoff of all its neighbors. Depending on which payoff is higher, the influence of either option can be tuned by means of a single parameter. We show that, in general, taking into account the environment promotes cooperation. Yet coveting the fitness of one's neighbors too strongly is not optimal. In fact, cooperation thrives best only if the influence of payoffs obtained in the traditional way is equal to that of the average payoff of the neighborhood. We present results for the prisoner's dilemma and the snowdrift game, for different levels of uncertainty governing the strategy adoption process, and for different neighborhood sizes. Our approach outlines a viable route to increased levels of cooperative behavior in structured populations, but one that requires a thoughtful implementation.
\end{abstract}


\end{frontmatter}

\section{Introduction}
\label{introduction}

Understanding the evolution of cooperation among unrelated individuals represents one of the major challenges of evolutionary biology and of behavioral sciences \citep{nowak_s06}. According to the principles of Darwinian selection, any behavior that brings benefits to others but not directly to oneself will soon disappear \citep{Darwin}. However, this is not fully consistent with observations that attest to the existence of cooperative behavior, with examples ranging from the communities of microorganisms to animal and human societies \citep{milinski_n87, binmore_94, colman_95, doebeli_el05}. In order to explain the emergence and maintenance of cooperation, evolutionary games, with the focus on social dilemmas, have provided several fundamental insights \citep{hofbauer_98, ohtsuki_n06, ren_cm06b, ohtsuki_prslb06}. And especially the prisoner's dilemma games and its extensions have been considered and studied frequently \citep{gibbons_jmb91, nowak_n92b, gibbons_bmb92, milinski_pnas98, moyano_jtb09, souza_mo_jtb09, chen_xj_pre09b, traulsen_pnas10, jimenez_jtb08, poncela_njp07, sysiaho_epjb05, hauert_n04, santos_prl05} in order to shed light on how cooperation can evolve and how it can be maintained. In its general form the prisoner's dilemma game states that the players must choose either cooperation or defection without knowing the decision of their co-players. A cooperator receives the reward $R$ when meeting another cooperator, but only the sucker's payoff $S$ when facing a defector. On the contrary, a defector exploiting the cooperator gets the temptation $T$, but only the punishment $P$ if encountering another defector. Because the above payoffs strictly satisfy the ranking $T>R>P>S$ and $2R>(T+S)$, eventually the defectors will prevail irrespective of what their opponent choose, and thus will become the dominant strategy. Altogether, we are faced with a social dilemma that if left ``untreated'' will lead to the tragedy of the commons \citep{hardin_g_s68}.

Over the past decades, several mechanisms have been identified that can offset an unfavorable outcome of social dilemmas and lead to the evolution of cooperation \citep{nowak_s06}. Examples include kin selection \citep{hamilton_wd_jtb64a}, direct and indirect reciprocity \citep{ohtsuki_jtb04, nowak_jtb98, nowak_n98, panchanathan_n04, ohtsuki_jtb04b, ohtsuki_jtb06b}, effective strategies such as the tit-for-tat \citep{imhof_jtb07, baek_pre08} or win-stay-lose-shift \citep{nowak_n93, chen_xj_pa08}, voluntary participation \citep{szabo_pre02d}, and of course spatially structured populations \citep{nowak_n92b, nowak_ijbc94, nakamaru_jtb97, nakamaru_jtb98}. Mostly notably, if players are arranged on a lattice and interact only with their nearest neighbors, then cooperators can survive by means of forming compact clusters which minimizes the exploitation by defectors and protects those cooperators that are located in the interior of such clusters \citep{nowak_n92b}. Along this line of research studies on the evolution of cooperation have received a substantial boost. For example, complex networks with the connectivity structure similar to that of social networks have been recognized as very beneficial for the evolution of cooperation \citep{abramson_pre01, santos_prl05, santos_jeb06, santos_pnas06, tang_epjb06, ohtsuki_n06, floria_pre09, gomez-gardenes_jtb08, rong_pre07, poncela_njp07, kuperman_epjb08, gomez-gardenes_prl07, du_wb_epl09}. In particular, the heterogeneity, or diversity, allows for cooperative behavior to prevail even if the temptations to defect are large \citep{szolnoki_epl07, perc_pre08, santos_n08, perc_pone10}. The mobility of players can also lead to an outbreak of cooperation, even when the conditions are noisy and do not necessarily favor the spreading of cooperators \citep{helbing_acs08, helbing_pnas09, jiang_ll_pre10}. Uncertainty, if appropriately tuned, may also have a positive impact on the evolution of cooperation \citep{perc_njp06a, vukov_pre06}. Moreover, there exist comprehensive reviews that capture succinctly recent advances on this topic \citep{szabo_pr07, perc_bs10, roca_plr09}.

However, while some of the works focus predominantly on the effects of individual properties, others build on the influence of external factors. Notably though, the conceptual relatedness of these seemingly very disparate mechanisms is often neglected. Here our aim is to propose an approach that integrates seamlessly between individual and external factors by means of a single parameter. The definition of fitness has already been modified for this purpose, for example based on the extension of Hamilton's rule \citep{lehmann_jeb06, doebeli_jeb06}, and here we also focus on this particular aspect of evolutionary games. As suggested in many previous works concerning also complex networks and processes taking place on them \citep{albert_rmp02, albert_prl01}, taking into account the fact that different nodes (players) have a different ability to compete successfully for a dominant position within the network is achieved best by assigning a fitness to each individual. Naturally, here we also consider individual fitness as being representative for the ability or potential of each individual to survive and reproduce. Moreover, we build on the fact that individual success in general depends on the inheritance as well as on environmental factors, and indeed many paradigmatic examples can been found in the biological and social sciences supporting this assertion \citep{Krakauer_nature05, english_be06, keller_trend97, thomas_78, rodrigues_jtb09}. For example, a young lion not only inherently knows how to suckle on its mother, but it has to gradually learn also how to prey and protect its territory according to the numbers of competing opponents. If it fails at either of these tasks, its chances of survival are slim. By considering the traditional payoff accumulation (what the players obtain upon playing with their neighbors) as something related to inheritance, and by considering the average payoff of all the neighbors as being representative for the environment, we propose a simple single-parameter dependent payoff function that allows us to determine just how much it pays to prefer one or the other, i.e., inheritance or the environment. In addition, the proposed payoff function incorporates a coevolutionary ingredient in that the influence of the two factors depends dynamically on its expected performance.

We focus on the prisoner's dilemma game, but present also detailed results for the snowdrift game. As the interaction network, we consider the square lattice with different numbers of neighbors in order to relevantly assess the importance of neighborhood size. We also examine the effects of different levels of uncertainty by strategy adoptions on the evolution of cooperation. Depending on the value of the parameter that determines how strongly individuals covet their neighbors (in the sense of wanting to rely completely on the average payoff of their neighborhood rather than on the traditionally obtained payoffs), we demonstrate that cooperation can be promoted substantially if compared to the traditional version of the game \citep{szabo_pre98, szabo_pre05}. Importantly though, we find that the facilitation of cooperation is optimal only if the inheritance and the environment are represented equally strong in the final fitness of each player. Since our findings are robust to variations of the governing evolutionary game, the neighborhood size, as well as to variations of the level of uncertainty governing the strategy adoptions, we conclude that the proposed approach outlines a viable route to resolving social dilemmas.

The paper is structured as follows. Section 2 features the methods and the description of evolutionary games, while section 3 contains the results. In the last section we summarize our conclusions.

\section{Methods}
\label{methods}

For simplicity, but without loss of generality, we consider variants of the prisoner's dilemma and the snowdrift game of which the outcomes depend on a single parameter only. For the prisoner's dilemma game, the payoffs are $T=b$, $R=1$ and $P=S=0$, where $1 \leq b \leq 2$ quantifies the temptation to defect and represent the advantage of defectors over cooperators. Although being in effect the so-called weak prisoner's dilemma in that $P=S$ rather than $P>S$, this version captures all the relevant aspects of the game \citep{nowak_n92b}. In order to test the validity of our conclusions, we also employ the snowdrift game with the payoffs $T=1+r$, $R=1$, $S=1-r$ and $P=0$, thus satisfying the ranking $T>R>S>P$, where $0 \leq r \leq 1$ represents the so-called cost-to-benefit ratio. Indeed, the snowdrift game is frequently studied as an alternative to the perhaps better known prisoner's dilemma \citep{hauert_n04, du_wb_epl09, wang_wx_pre06}.

As the interaction network, we use $L \times L$ square lattices with periodic boundary conditions. Each vertex $i$ is initially designated as a cooperator ($s_{i}=C$) or defector ($s_{i}=D$) with equal probability. The game is iterated forward in accordance with the Monte Carlo simulation procedure comprising the following elementary steps. First, player $i$ acquires its payoff $P_{i}$ by playing the game with all its neighbors. Next, the environment of player $i$ is assessed by the average payoff of all its neighbors $\overline{P}$, that is,
\begin{equation}
\overline{P}=\frac{ {\bf \sum}_{j=1}^k P_{j}}{k},
\end{equation}
where $k$ denotes the neighborhood size of player $i$, $P_{j}$ represents the payoff of player $j$ who is one of the neighbors of player $i$, and the sum runs over all the neighbors of player $i$.

Before proceeding with the details of how individual fitness is determined, we would like to motivate our approach better, in particular describing why inheritance and environment are represented by individual (traditional) payoffs and the average payoff of all the neighbors, respectively. From the biological point of view, inheritance refers to the fact that individuals pass down their genetic material to their offspring. In the context of evolutionary games, this corresponds to players passing their strategy to the next generation based on their payoffs \citep{szabo_pre98, ohtsuki_jtb06}. Naturally, each accumulated payoff at present is the best reflection of the strategy which was inherited from the previous generation. On the other hand, in social systems the performance of each individual is affected not just by inheritance, but also by environmental factors \citep{rodrigues_jtb09, chalambor_funec07, Strassmann89}, implying that to some extent individual success is related to the performance of its neighbors or rather the neighborhood as a whole. In order to capture this influence succinctly, we consider the average payoff of all the neighbors as the simplest measure to assess the influence of the environment. Motivated by the fact that the environment (here represented by $\overline{P}$) may be just as important as inheritance (here represented by $P_i$), but also by the fact that in general the impact of these two factors may vary, we finally evaluate the fitness of player $i$ according to
\begin{equation}
{\it f_{i}} = \left\{ \begin{array}{ll}
(0.5-u) \times \overline{P}+ (0.5+u) \times P_{i} & \textrm{if ($P_{i}>\overline{P}$)},\\
\\
(0.5+u) \times \overline{P}+ (0.5-u) \times P_{i} & \textrm{if ($P_{i}<\overline{P}$)},\\
\\
0.5 \times \overline{P}+ 0.5 \times P_{i} & \textrm{if ($P_{i}=\overline{P}$)},
\end{array} \right.
\end{equation}
where the selection parameter $0 \leq u \leq 0.5$ is used for fine-tuning. Evidently, for $u=0$ both influences determine the final fitness of player $i$ in equal capacity. For $u>0$, however, the better performing influence will be preferred, i.e., represented stronger in the final fitness. In the limit case of $u=0.5$ the fitness $f_{i}$ is absolutely determined either by the environment or by inheritance, whichever is performing better at the time. Alternatively, Eq.~(2) can also be interpreted as follows: Before each generation (during the simulation, each full Monte Carlo step is regarded as a new generation), we assume that the influence of inheritance and environment on individual development is the same because we cannot objectively predict the magnitude of their influence before the appearance of a new generation. However, after the impact of both is evaluated, the influence will change accordingly. If the performance of neighbors is better, the player may benefit from the environment. Otherwise, the influence of its neighbors may be reduced or is kept constant. Following the determination of fitness, player $i$ adopts the strategy $s_{j}$ from its randomly selected neighbor $j$ (whose fitness $f_{j}$ is determined in the same way as $f_{i}$) via the probability
\begin{equation}
W(s_j \rightarrow s_i)=\frac{1}{1+\exp[(f_i-f_j)/K]},
\end{equation}
where $K$ denotes the amplitude of noise or its inverse ($1/K$) the so-called intensity of selection \citep{szabo_pre98}. Positive values of $K$ imply that better performing players are readily imitated, but it is not impossible to adopt the strategy of a player performing worse. Such errors in judgment can be attributed to mistakes and external influences that affect the evaluation of the opponent. During a full Monte Carlo step (MCS) all players will have a chance to pass their strategy once on average.

Results of Monte Carlo simulations presented below were obtained on populations comprising up to $400 \times 400$ individuals, whereby the fraction of cooperators $F_{c}$ was determined within $10^5$ full MCS after sufficiently long transients were discarded. Moreover, final results were averaged over up to $40$ independent runs for each set of parameter values in order to assure suitable accuracy.

\section{Results}
\label{results}

\begin{figure}
\begin{center}
\includegraphics[width=7.5cm]{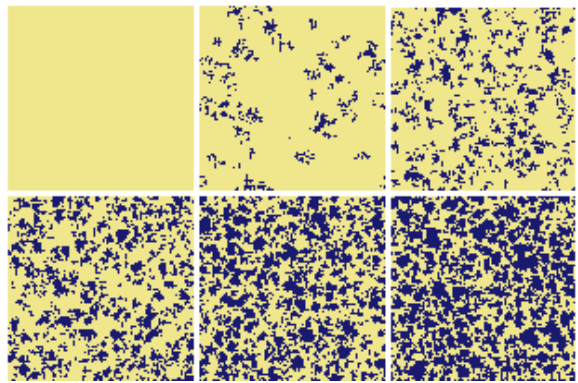}
\caption{Characteristic distributions of cooperators (blue) and defectors (yellow) for different values of the parameter $u$. From top left to bottom right $u = 0.5$, $0.4$, $0.3$, $0.2$, $0.1$ and $0$, respectively. All panels depict results obtained for $b=1.45$ and $K=0.1$ on a $100 \times 100$ square lattice.}
\label{fig1}
\end{center}
\end{figure}

As is known, in the prisoner's dilemma game the cooperators will be decimated fast even if the temptations to defect are moderate. It is thus challenging to identify non-trivial mechanisms that may sustain cooperation under such conditions. In order to address this puzzle, we consider first the effect of the redefined fitness, as given by Eq.~(2). Figure~\ref{fig1} shows the characteristic spatial distributions of cooperators and defectors for different values of the parameter $u$. If $u=0.5$ (top left panel), where each player's performance is absolutely determined by either the inheritance or the environment (depending on performance), cooperators will go extinct, the final outcome thus being complete dominance of defectors. However, upon a slight decrease of $u$, the survival of cooperators becomes viable in that a small fraction of cooperators can prevail by means of forming small clusters or patches on the spatial grid. By continuing to decrease $u$, the clusters of cooperators become larger and more common, which ultimately results in averting the impeding social decline. More interestingly, for $u=0$ (bottom right panel), when the influence of inheritance is equal to that of the environment, cooperators thrive best, and may even outperform defectors. Hence, these results suggest that the parameter $u$, determining the composition of the fitness of each player can substantially promote cooperation, enabling its maintenance where otherwise defection would reign completely. Yet coveting the fitness of one's neighbor too strongly (which is implied by $u=0.5$), even if at the moment the neighbors are performing much better, is not optimal for the evolution of cooperation.

\begin{figure}[ht]
\begin{center}
\includegraphics[width=7.5cm]{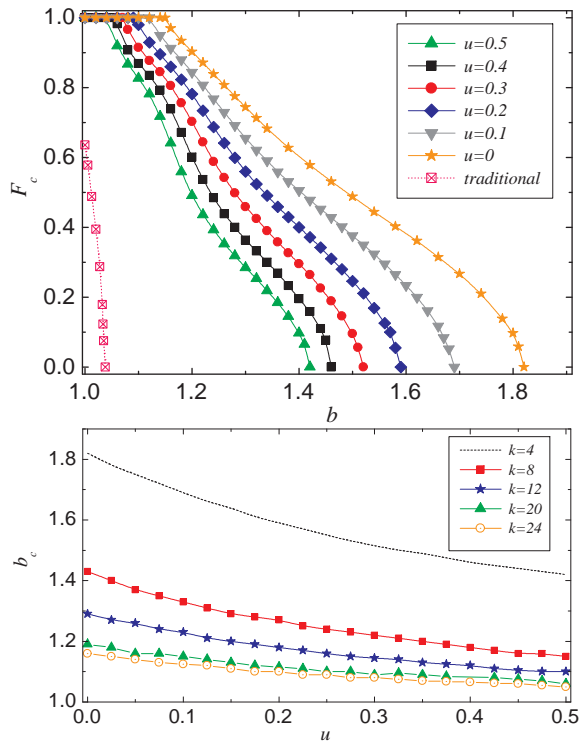}
\caption{Top panel: Frequency of cooperators $F_c$ in dependence on the parameter $b$ for different values of the selection parameter $u$. From left to right $u$ = 0.5, 0.4, 0.3, 0.2, 0.1 and 0, respectively (the outcome of the traditional version of the game is denoted dashed). Bottom panel: Critical threshold values $b = b_c$, marking the transition to the pure $D$ phase (extinction of cooperators), in dependence on the selection parameter $u$ for different neighborhood sizes. If compared to the traditional version of the game (both panels), it can be observed that cooperation can be maintained by significantly higher values of $b$, and moreover, that larger neighborhood sizes may lessen the promotive impact significantly. Depicted results in both panels were obtained for $K= 0.1$.}
\label{fig2}
\end{center}
\end{figure}

In order to provide a quantitative assessment of the impact of different values of $u$, we show in Fig.~2 how the fraction of cooperators $F_{c}$ and the critical temptation to defect $b_c$, at which cooperators go extinct, depends on this newly introduced parameter. Results presented in the top panel of Fig.~\ref{fig2} depict $F_c$ in dependence on the parameter $b$ for different values of $u$. One can find, compared with the traditional version of the game, that the introduction of $u$ can substantially promote the emergence and maintenance of cooperation. Moreover, the presented results demonstrate explicitly that the switch of the parameter $u$ from $0.5$ to $0$ makes cooperators stronger and more resilient to the invasion of defectors. These quantitative results clearly attest to the fact that the environment plays a vital role in individual development, specifically by the evolution of cooperation, yet redundantly leaning on it (or the traditional accumulation of payoffs), which is implied by $u=0.5$, will not be optimal.

\begin{figure}
\begin{center}
\includegraphics[width=6.7cm]{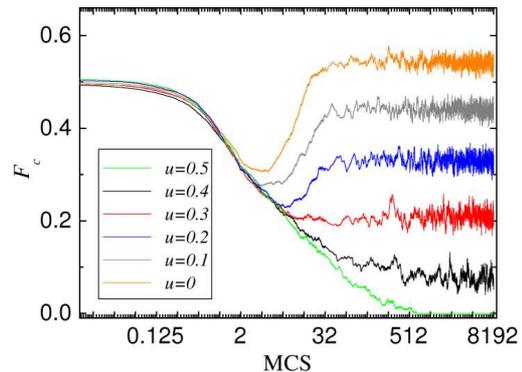}
\caption{Time courses depicting the evolution of cooperation for different values of $u$. All time courses were obtained as averages over $20$ independent realizations for $b=1.45$ and $K=0.1$ on a $200 \times 200$ square lattice. Note that the horizontal axis is logarithmic and that values of ${\it F_c}$ were recorded also in between full Monte Carlo steps to ensure a proper resolution.}
\label{fig3}
\end{center}
\end{figure}

It's also interesting to consider how the critical threshold value $b_c$, marking the extinction of cooperators, varies in dependence on the selection parameter $u$ for different neighborhood sizes. From the bottom panel of Fig.~\ref{fig2}, it can be observed that the value of $b_c$ decreases monotonously from $1.82$ to $1.42$ while increasing $u$ from $0$ to $0.5$ in case of the traditional square lattice ($k=4$). However, if the neighborhood size on the square lattice is enlarged, this effect becomes less and less pronounced as $k$ increases, and in fact at $k=24$ only a marginal difference in $b_c$ can be observed if comparing the $u=0$ and the $u=0.5$ case. This result is in fact expected since increasing the neighborhood size will gradually lead to well-mixed conditions \citep{szabo_pre09}, but it also implies directly that the observed phenomenon is inherently routed in the spatiality of the interaction structure. Below we will provide further evidence supporting such a conclusion when we investigate how different values of $K$ affect the evolution of cooperation by different values of $u$. Nevertheless, it is also worth pointing out that the general features of our results remain intact upon changing the neighborhood size, which vouches for their robustness.

\begin{figure}[ht!]
\begin{center}
\includegraphics[width=7.5cm]{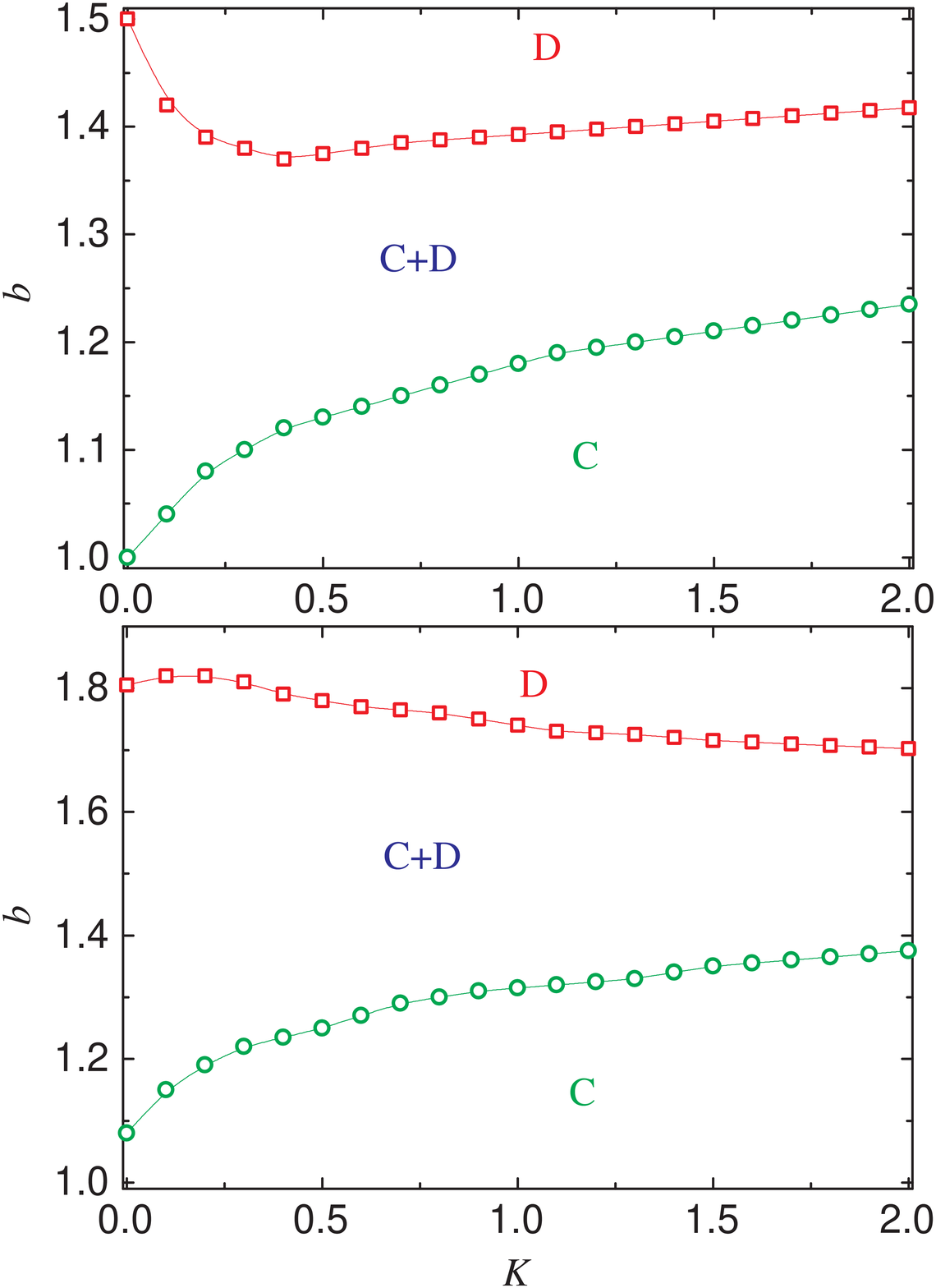}
\caption{Full $b-K$ phase diagrams for $u=0.5$ (top panel) and $u=0$ (bottom panel), obtained via Monte Carlo simulations of the prisoner's dilemma game on the square lattice. The green and red lines mark the border between stationary pure C and D phases and the mixed C+D phase, respectively. In contrast with previous works considering the square lattice \citep{vukov_pre06, szabo_pre05}, it can be observed that for $u=0.5$ (top panel) there exists an intermediate uncertainty in the strategy adoption process (an intermediate value of $K$) for which the survivability of cooperators is worst, i.e., $F_c$ is minimal rather than maximal. This suggests that the interaction topology is indirectly affected and may give rise to overlapping triangles if either the inheritance or the environment are favored too strongly depending on their relative performance at each particular moment in time. Conversely, while the borderline separating the pure C and the mixed C+D phase for the $u=0$ case (bottom panel) exhibits a qualitatively identical outlay as for the $u=0.5$ case, the D $\leftrightarrow$ C+D transition is qualitatively different. Note that in the bottom panel there indeed exists an intermediate value of $K$ for which $F_c$ is maximal rather than minimal, which is in agreement with what can be expected for interaction topologies that lack overlapping triangles.}
\label{fig4}
\end{center}
\end{figure}

In order to explain how and why different values of $u$ promote cooperation, we first examine time courses of $F_c$ for different values of the selection parameter $u$. From Fig.~\ref{fig3}, it becomes obvious fast that in the early stages of the evolutionary process (note that values of $F_c$ were recorded also in between full Monte Carlo steps) the performance of defectors is better than that of cooperators. This is in fact what one would expect, since defectors, as individuals, should be more successful than cooperators, which in turn should manifest in the decimation of the later. What is not necessarily expected, is that the tide shifts in favor of cooperators rather strongly following their initial decline, and in fact the more so the smaller the value of $u$. In particular, when the value of $u$ is large, i.e., close or equal to $0.5$, cooperators will ultimately go extinct or pend at the brink of extinction. With the decrease of $u$, however, the tide may change strongly in favor of the cooperators. For $u=0$, for example, it can be observed that the initial downfall of cooperators is rather shallow, and ultimately, they can restore their presence on the spatial grid in equal capacity as the defectors. This suggests that in the initial stages of the game, when the cooperators are not yet clustered, the defectors can successfully exploit them. However, as the cooperative clusters form, they become impervious to the defector attacks, which is due not only to spatial reciprocity, but also due to the newly identified mechanism which can amplify the effect of spatial reciprocity substantially. Ultimately, the cooperators can therefore survive at higher temptations to defect than would be possible by spatial reciprocity alone.

It is next of interest to examine the evolution of cooperation for different values of $u$ in dependence on the uncertainty by strategy adoptions. The latter can be tuned via $K$ in Eq.~(3), which acts as a temperature parameter in the employed Fermi strategy adoption function \citep{szabo_pre98}. Accordingly, when $K \to \infty$ all information is lost and the strategies are adopted by means of a coin toss. Figure~\ref{fig4} features full $b-K$ phase diagrams for the square lattice at $u=0.5$ (top) and $u=0$ (bottom). Interestingly, $u=0.5$ eradicates (as do interaction networks incorporating overlapping triangles \citep{szabo_pre05, szolnoki_pre09c}) the existence of an optimal $K$, as can be observed from the phase diagram presented in the top panel, which exhibits an inverted bell-shaped $D \leftrightarrow C+D$ transition line, indicating the existence of the worst ($K \approx 0.4$) rather than an optimal temperature for the evolution of cooperation. This in turn implies that introducing a strong preference towards either the inheritance (the fitness as determined by the traditional accumulation of payoffs) or the environment (the fitness as determined by the average payoff of all the neighbors) effectively alters the interaction network. While the square lattice obviously lacks overlapping triangles and thus enables the observation of an optimal $K$, trimming the importance via $u$ seems to effectively enhance linkage among essentially disconnected triplets and thus precludes the same observation. A similar phenomenon was observed recently in public goods games, where the joint membership in large groups was also found to alter the effective interaction network and thus the impact of uncertainly on the evolution of cooperation \citep{szolnoki_pre09c}. Conversely, the phase diagram presented in the bottom panel of Fig.~\ref{fig4} is well-known (at least qualitatively), implying the existence of an optimal level of uncertainty for the evolution of cooperation, as was previously reported in \citep{perc_njp06a, vukov_pre06}. In particular, note that the $D \leftrightarrow C+D$ transition line is bell shaped, indicating that $K \approx 0.15$ is the optimal temperature at which cooperators are able to survive at the highest value of $b$. This phenomenon can be interpreted as an evolutionary resonance \citep{perc_njp06a}, albeit it can only be observed on interaction topologies lacking overlapping triangles \citep{szabo_pre05, szolnoki_pre09c}. Altogether, these results confirm that the observed promotion of cooperation is routed strongly in the spatiality of the interaction network, which is clearly manifested by an extensive gap between the $C \leftrightarrow C+D$ and the $D \leftrightarrow C+D$ transition lines at $u=0$, indicating that cooperators may survive even if $b$ is close to the maximal value.

\begin{figure}[ht!]
\begin{center}
\includegraphics[width=7.5cm]{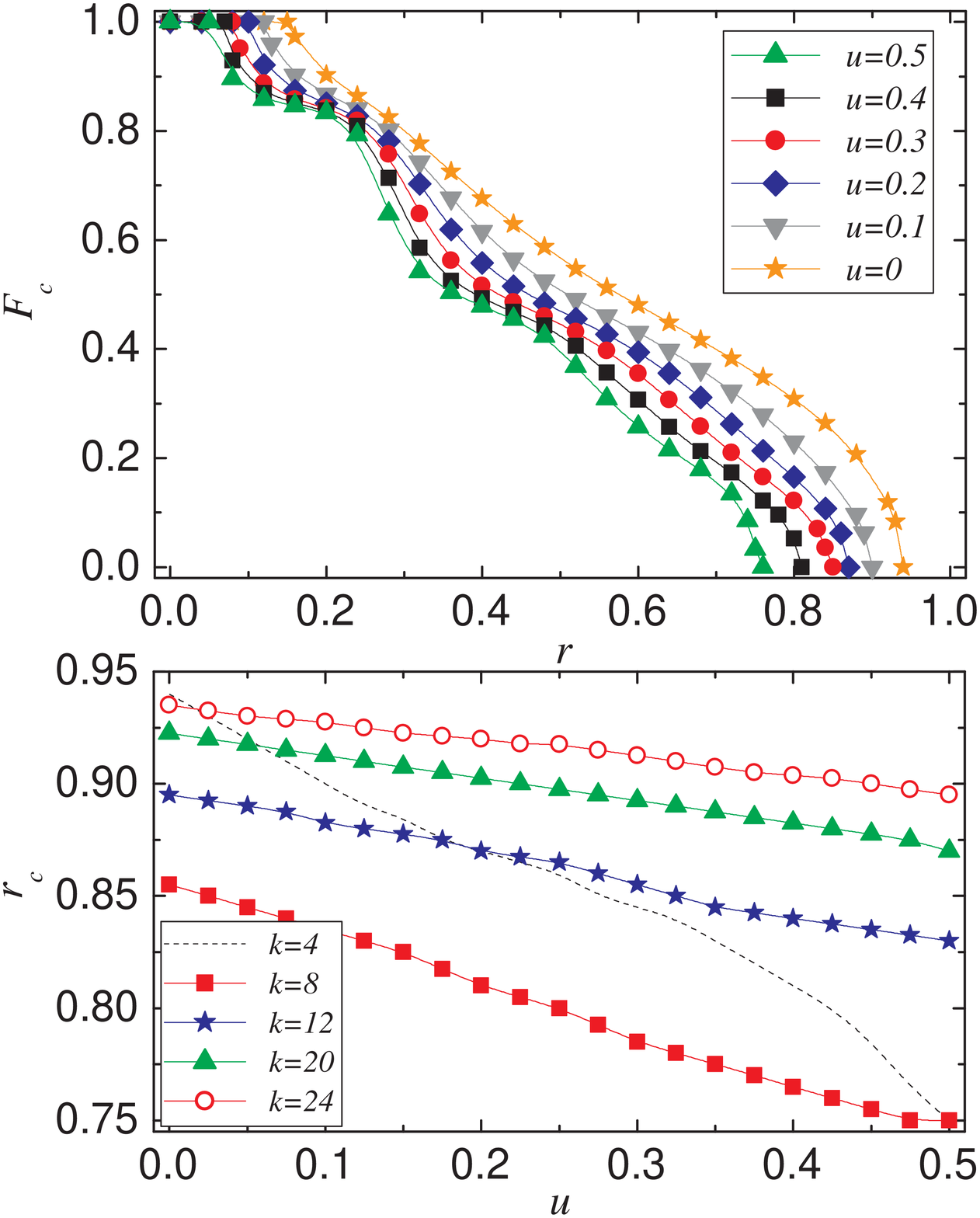}
\caption{Top panel: Frequency of cooperators $F_c$ in dependence on the parameter $r$ for different values of the selection parameter $u$. From left to right $u$ = 0.5, 0.4, 0.3, 0.2, 0.1 and 0, respectively. Bottom panel: Critical threshold values of the parameter $r = r_c$, marking the transition to the pure $D$ phase (extinction of cooperators), in dependence on the selection parameter $u$ for different neighborhood sizes. It is to be emphasized that these results are qualitatively in agreement with those obtained for the prisoner's dilemma game (see Fig.~\ref{fig2}). Depicted results in both panels were obtained for $K= 0.1$.}
\label{fig5}
\end{center}
\end{figure}

Finally, it is of interest to explore the generality of our observations by means of different evolutionary games. Due to the famous claim that the spatial structure may inhibit the evolution of cooperation in the snowdrift game \citep{hauert_n04}, the snowdrift game naturally becomes an appropriate candidate for this task. Figure~\ref{fig5} depicts the fraction of cooperators $F_c$ in dependence on the parameter $r$ for different values of $u$. Similarly as in Fig~\ref{fig2}, it can be observed that with the value of $u$ decreasing, the evolution of cooperation is facilitated, which is qualitatively consistent with the results obtained for the prisoner's dilemma game. Interestingly though, the effect is less pronounced, which may be attributed to the fact that the spatiality is indeed less crucial (is in fact detrimental) for the evolution of cooperation in the snowdrift game, than it is for the evolution of cooperation in the prisoner's dilemma. This assertion if fully confirmed upon examining the dependence of the critical $r=r_c$ for different neighborhood sizes $k$. We remind the reader that for the prisoner's dilemma game the fact that larger values of $k$ decrease the level of cooperation is expected since increasing the neighborhood size will gradually lead to well-mixed conditions. Since the spatial structure is known to be crucial for the sustenance of cooperators in the prisoner's dilemma game \citep{nowak_n92b}, this is an expected result that is not difficult to understand. It also means that the spatiality (the fact that interactions are limited to neighbors on the lattice) is crucial for the observed promotion of cooperation. The results for the snowdrift game presented in the bottom panel of Fig~\ref{fig5} are different. The paper by \citet{hauert_n04} identified key differences in the pattern formation of cooperators by the snowdrift game that is due to the different payoff structure (if compared to the prisoner's dilemma game). While in the spatial prisoner's dilemma cooperators can survive by forming large, compact clusters, in the spatial snowdrift game cooperators form only small filament-like clusters. The latter make it advantageous to adopt strategies that are opposite to neighboring strategies, ultimately resulting in the fact that the spatial structure actually inhibits the evolution of cooperation in the snowdrift game. Our results in the bottom panel of Fig~\ref{fig5} agree with this in that larger values of $k$ (larger neighborhoods), decreasing the impact of spatiality, promote cooperation in the snowdrift game (note that values of $r_c$ become higher for larger $k$). Thus, the impact of $k$ is opposite to that for the prisoner's dilemma game, which is in agreement with the argumentation proposed by \citet{hauert_n04}. On the other hand, the impact of the parameter $u$ is the same in that the smaller it is the larger the value of $F_c$. A special case is the result for $k=4$ by the snowdrift game, where the parameter $u$ seems to play an even more crucial role than for higher values of $k$. A precise reason for this was impossible for us to find. Intuitively, for $k=4$ the conflict between the fact that spatial structure inhibits the evolution of cooperation while small values of $u$ promote it is expressed most severely, thus leading to the strong dependence, i.e., much stronger than for larger values of $k$ or for any value of $k$ in the prisoner's dilemma game. Note that in the latter game the aforementioned conflict does not emerge because there the spatial structure at $k=4$ is in fact optimal for the evolution of cooperation, while for the snowdrift game it is the most prohibitive. Nevertheless, these results support the fact that the newly identified mechanism that boosts the effect of spatial reciprocity is generally valid, and should thus be observable also under circumstances that were not explicitly taken into account in this paper.

\section{Discussion}
\label{discussion}

The evolutionary success of cooperators in social dilemmas is an important and vibrant topic. In order to provide insights into this fascinating phenomenon, the prisoner's dilemma, as a basic and general metaphor for the problem, is commonly employed. In its original form, it is to be expected that rational individuals will favor defection of cooperation. This can be averted by introducing spatially structured interactions \citep{nowak_n92b}. In the spatial setting, cooperators are able to survive by forming compact clusters, which disables the defectors to exploit those that are located in the interior of such clusters. However, if the temptation to defect is sufficiently large, the spatial reciprocity may fail to sustain cooperation. To overcome this, various additional mechanisms that may promote cooperation have been proposed. Some of them focused on individual properties of players, as for example the teaching activity \citep{szolnoki_epl07}, while others focused on the external factors (or the environment), as for example the structure of the interaction network \citep{abramson_pre01}. Motivated by this fact, and by the concept of fitness as often defined from the biological viewpoint, we introduce here an alternative definition of fitness that is composed from inheritance (the payoffs as obtained by playing the game with the neighbors) and the environment (the average payoff of all the neighbors). Depending on which payoff is higher, the influence of either option can be tuned by means of a single parameter $u$. Our approach is of course a minimalist one, allowing for proof of principle rather than accurate claims about specific setups, yet it demonstrates effectively that the concept of fitness is amenable to simple adjustments that may have wanted consequences for the evolution of cooperation. In particular, by means of systematic simulations, we have shown that considering the environment as a necessary composition of fitness can greatly promote the evolution of cooperation, especially if compared to the traditional version of the game (either the prisoner's dilemma or the snowdrift game) that does not take into account the role of the environment in individual development. But also, we demonstrate that if the individuals are too avid in coveting what their neighbors have (in terms of payoffs), the evolution of cooperation will not be optimally promoted. The best is to adjust both influences to be represented equally strong.

In addition, we have presented a detailed analysis of the promotion effect with the help of time courses and the outcome of the games by different levels of uncertainty governing the strategy adoptions. Although defection is prevalent in the early stages of the evolutionary process, small values of the parameter $u$ can revert this trend, typically so that the few remaining cooperators form very compact clusters that are impervious to defector attacks. These clusters, although initially small and rare, may inflate fast and ultimately outperform the defectors. Also interesting is the fact that the introduction of $u$ seems to alter the effective interaction topology of the square lattice. If the value of $u$ is large, i.e., if the average payoff of the neighbors is considered as too strong a factor in the determination of individual fitness, there exists only the ``worst level'' of uncertainty, at which cooperators go extinct by the smallest temptation to defect. Conversely, if $u=0$, which constitutes the optimal setup for the evolution of cooperation, there exists an optimal level of uncertainty, which can only be observed if the interaction topology is lacking overlapping triangles \citep{szabo_pre05}. However, since the actual topology always remains unchanged, we attribute the effect on the evolution of cooperation to the possible alteration of the effective interaction topology by means of previously unrelated individuals due to the consideration of environmental factors.

Lastly, to test whether our approach is effective also in evolutionary games other than the prisoner's dilemma, we explore the evolution of cooperation in the snowdrift game. We obtain qualitatively identical results as by the prisoner's dilemma game, with some minor differences existing with regards to the impact of different neighborhood sizes. Nevertheless, the conclusion that cooperation thrives best only if the influence of payoffs obtained in the traditional way is equal to that of the average payoff of the neighborhood remains valid, thus constituting a viable route to increased levels of cooperation in structured populations.

\section*{Acknowledgments}
ZW acknowledges support from the Center for Asia Studies of Nankai University (Grant No. 2010-5) and from the National Natural Science Foundation of China (Grant No. 10672081). WD acknowledges support from the Innovative Research Groups of the National Natural Science Foundation of China (Grant No. 60921001). ZR acknowledges support from the National Natural Science Foundation of China (Grant No. 61004098). MP acknowledges support from the Slovenian Research Agency (grant Z1-2032).


\end{document}